\documentclass{PoS}

\usepackage[utf8]{inputenc}
\usepackage[T1]{fontenc}
\usepackage{amsmath,amsthm, amssymb, latexsym}
\usepackage{slashed}

\usepackage{latexsym}
\usepackage{subfigure}
\usepackage{multicol}
\usepackage{multirow}
\usepackage{mathrsfs}
\usepackage{verbatim}
\usepackage{latexsym}
\usepackage{subfigure}
\usepackage{multicol}

\usepackage[absolute,overlay]{textpos}

\title{Pion and Kaon Distribution Amplitudes from lattice QCD: towards the continuum limit}

\ShortTitle{Pion and Kaon Distribution Amplitudes from lattice QCD: towards the continuum limit}

\author{Gunnar S. Bali, Vladimir M. Braun, Meinulf Göckeler, Michael Gruber, Fabian Hutzler,\ \speaker{Piotr Korcyl}\thanks{Present address: M. Smoluchowski Institute of Physics, Jagiellonian University, ul. \L ojasiewicza 11, 30-348 Krak\'ow, Poland}, \ Andreas Schäfer, Philipp Wein \\
        Institut f\"ur Theoretische Physik, Universit\"at Regensburg, 93040 Regensburg, Germany\\
        E-mail: \email{piotr.korcyl@uj.edu.pl}}

\abstract{
We present the current status of a non-perturbative lattice calculation of the moments of the pion and kaon distribution amplitudes by the RQCD collaboration. Our investigation is carried out using $N_f=2+1$ dynamical, non-perturbatively O(a)-improved Wilson fermions on the CLS ensembles with 5 different lattice spacings and pion masses down to the physical pion mass. A combined continuum and chiral extrapolation to the physical point is performed along two independent quark mass trajectories simultaneously. We employ momentum smearing in order to decrease the contamination by excited states and increase statistical precision.
}

\FullConference{The 36th Annual International Symposium on Lattice Field Theory - LATTICE2018\\
		22-28 July, 2018\\
		Michigan State University, East Lansing, Michigan, USA.}

\begin{document}

\section{Introduction}

Pion and kaon distribution amplitudes (DA) are examples of non-perturbative objects describing the internal structure of hadrons. The importance of such functions in QCD comes from factorization theorems which allow to disentangle a generic cross-section into a process specific hard part calculable within a perturbative framework and a process-independent non-perturbative part. The latter can be either extracted from experimental data if such is available or estimated numerically using lattice QCD techniques. The pion DA which is investigated in this study describes the quark-antiquark momentum composition of the lowest Fock state with only a single valence quark-antiquark pair. More precisely it is the quantum amplitude that the pion moving with longitudinal momentum $p$ is built of a pair of quark and antiquark moving with momentum $x p$ and $(1-x)p$ respectively. Defined as it is, it uses a light-cone formulation of QCD, and hence is not directly amenable to numerical studies on euclidean lattices. A traditional way to circumvent this limitation is to study its moments which can be expressed as matrix elements of local operators with derivatives between a vacuum and a pion state. So far most of the attention was devoted to the second moment of the pion DA \cite{PhysRevD.74.074501, PhysRevD.83.074505, PhysRevD.92.014504}. In this contribution we summarize our ongoing efforts aiming at obtaining a first continuum estimate of that moment. We build upon our earlier experience \cite{PhysRevD.92.014504} improved by the use of momentum smearing \cite{Bali:2016lva} which we have investigated in application to the case of pion DA moments in Ref.\cite{Bali:2017ude}.

This contribution has the following structure. In the next section we briefly recall the formulation that we employ to estimate the second moment of the pseudoscalar meson DA. In section \ref{sec. lattice} we describe the gauge configuration ensembles analyzed so far. Next, we discuss some details of our current strategy to extract estimates of the pion and kaon DA second moment in section \ref{sec. improvements}. We present our chiral and continuum extrapolation formulae in section \ref{sec. extrapolations} as well as some preliminary results for some of the extrapolated quantities. We conclude in section \ref{sec. conclusions}.

\section{Pseudoscalar meson distribution amplitudes}
\label{sec. pion da}

In this work we consider the pion and kaon distribution amplitudes. In particular, the pion DA is defined through the following light-like correlation
\begin{equation}
\langle 0 | \bar{d}(z_2 n) \slashed n \gamma_5 [z_2 n, z_1 n] u(z_1 n)| \pi(p) \rangle =  
i f_{\pi}(p \cdot n) \int_0^1 dx e^{-i(z_1x+z_2(1-x)) p \cdot n} \phi_{\pi}(x,\mu^2). 
\label{eq. light-like correlation}
\end{equation}
Neglecting isospin breaking effects leads to a function 
$\phi_{\pi}(x)$ which is symmetric under the interchange of momentum fraction 
$x \rightarrow (1-x)$,
\begin{equation}
\phi_{\pi}(x, \mu^2) = \phi_{\pi}(1-x,\mu^2).
\end{equation}
Therefore, only the even moments of the momentum fraction difference,  $\xi = x - ( 1- x)$,
\begin{equation}
\langle \xi^{2n} \rangle = \int_0^1 dx (2x - 1)^{2n} \phi_{\pi}(x,\mu^2),
\end{equation}
 which can be estimated on the lattice, are different from zero. Alternatively, using approximate conformal symmetry, one can re-parametrize the pion DA in terms of Gegenbauer polynomials with coefficients $a_{2n}^{\pi}$, 
\begin{equation}
\phi_{\pi}(x,\mu^2) = 6x(1-x)\Big[ 1+ \sum_n a^{\pi}_{2n}(\mu) C^{3/2}_{2n}(2x-1)\Big].
\end{equation}
In this study we are interested in the second moments, either $\langle \xi^2 \rangle$ or $a_{2}$. Both are renormalization scheme dependent and therefore we quote their $\overline{\textrm{MS}}$ values at the renormalization scale of 2 GeV.

In order to express the non-local, light-like operators of Eq.~\eqref{eq. light-like correlation}, one Taylor expands them and obtains local operators with derivatives. The general formulae were discussed in Ref. \cite{PhysRevD.92.014504}, here we just recall the two lattice operators which are relevant for the second moment of the pion DA (round brackets meaning symmetrization of indices and trace subtraction),
\begin{equation}
\mathcal{O}^{\pm}_{\rho \mu \nu}(x) = \bar{d}(x) \Big[ \overleftarrow{D}_{(\mu} \overleftarrow{D}_{\nu}
\pm 2 \overleftarrow{D}_{(\mu} \overrightarrow{D}_{\nu} 
+ \overrightarrow{D}_{(\mu} \overrightarrow{D}_{\nu} \Big] \gamma_{\rho)}\gamma_5 u(x). \end{equation}
Their bare matrix elements between the vacuum and a pion state are, up to renormalization effects, proportional to
\begin{align}
\langle 0 | \mathcal{O}^{-}_{\rho \mu \nu}(x) | \pi \rangle & \sim \langle [ x - (1 -x ) ]^2 \rangle = \langle \xi^2 \rangle, \\
\langle 0 | \mathcal{O}^{+}_{\rho \mu \nu}(x) | \pi \rangle & \sim \langle [ x + (1 -x ) ]^2 \rangle = \langle 1^2 \rangle. 
\label{eq. operators}
\end{align}
In order to extract the above matrix elements from a lattice simulation we estimate two kinds of two-point correlation functions
\begin{align}
C_{\rho}(t,\mathbf{p}) &= a^3 \sum_{\mathbf{x}} e^{-i \mathbf{p} \mathbf{x}} \langle \mathcal{O}_{\rho}(\mathbf{x},t)
J_{\gamma_5}(0) \rangle, \\
C^{\pm}_{\rho \mu \nu}(t,\mathbf{p}) &= a^3 \sum_{\mathbf{x}} e^{-i \mathbf{p} \mathbf{x}} \langle \mathcal{O}^{\pm}_{\rho \mu \nu}(\mathbf{x},t)
J_{\gamma_5}(0) \rangle,
\label{eq. correlation functions}
\end{align}
which we use to construct the following ratios
\begin{equation}
R^{\pm}_{\rho \mu \nu, \sigma}(t, \mathbf{p}) = \frac{C^{\pm}_{\rho \mu \nu}(t,\mathbf{p})}{C_{\sigma}(t,\mathbf{p})} \sim \frac{p_{\rho} p_{\mu} p_{\nu}}{p_{\sigma}} R^{\pm}.
\label{eq. ratio}
\end{equation}
By construction $R^{\pm}_{\rho \mu \nu, \sigma}(t, \mathbf{p})$ should exhibit plateaux for large $t$ separations from which the values $R^{\pm}$ can be extracted by fitting. Eventually, $R^{\pm}$ can be combined with renormalization constants in order to provide estimates of $\langle \xi^2 \rangle$ and $a_2$,
\begin{align}
\langle \xi^2 \rangle^{\overline{\textrm{MS}}} &= \zeta_{11} R^- + \zeta_{12} R^+, \\
a_2^{\overline{\textrm{MS}}} &= \frac{7}{12} \Big[ 5 \zeta_{11} R^- + (5\zeta_{12} - \zeta_{22} ) R^+ \Big].
\end{align}
The $\zeta_{ij}$ renormalization constants are estimated non-perturbatively
in the RI'/SMOM scheme and matched to the $\overline{\textrm{MS}}$ scheme at NLO \cite{PhysRevD.92.014504}.

\section{CLS ensembles}
\label{sec. lattice}

\begin{figure}
\begin{center}
\includegraphics[width=0.3\textwidth]{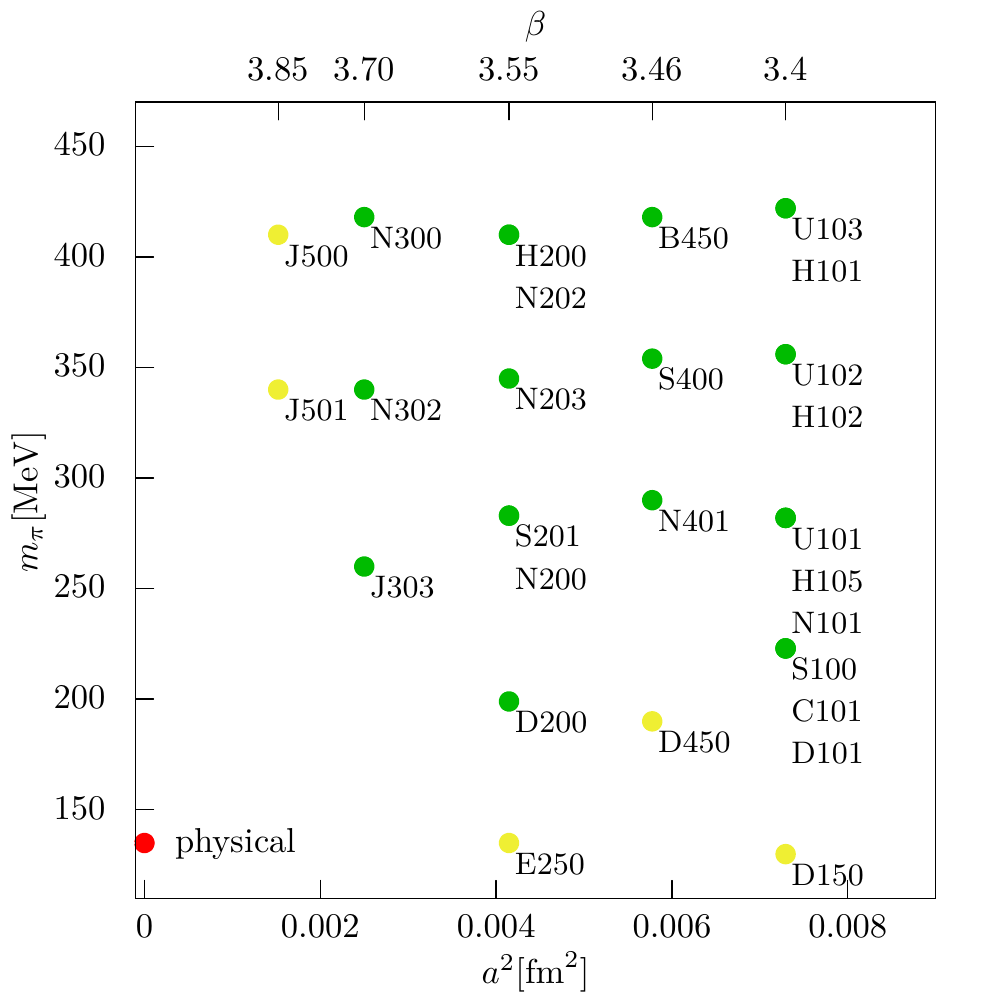}
\includegraphics[width=0.3\textwidth]{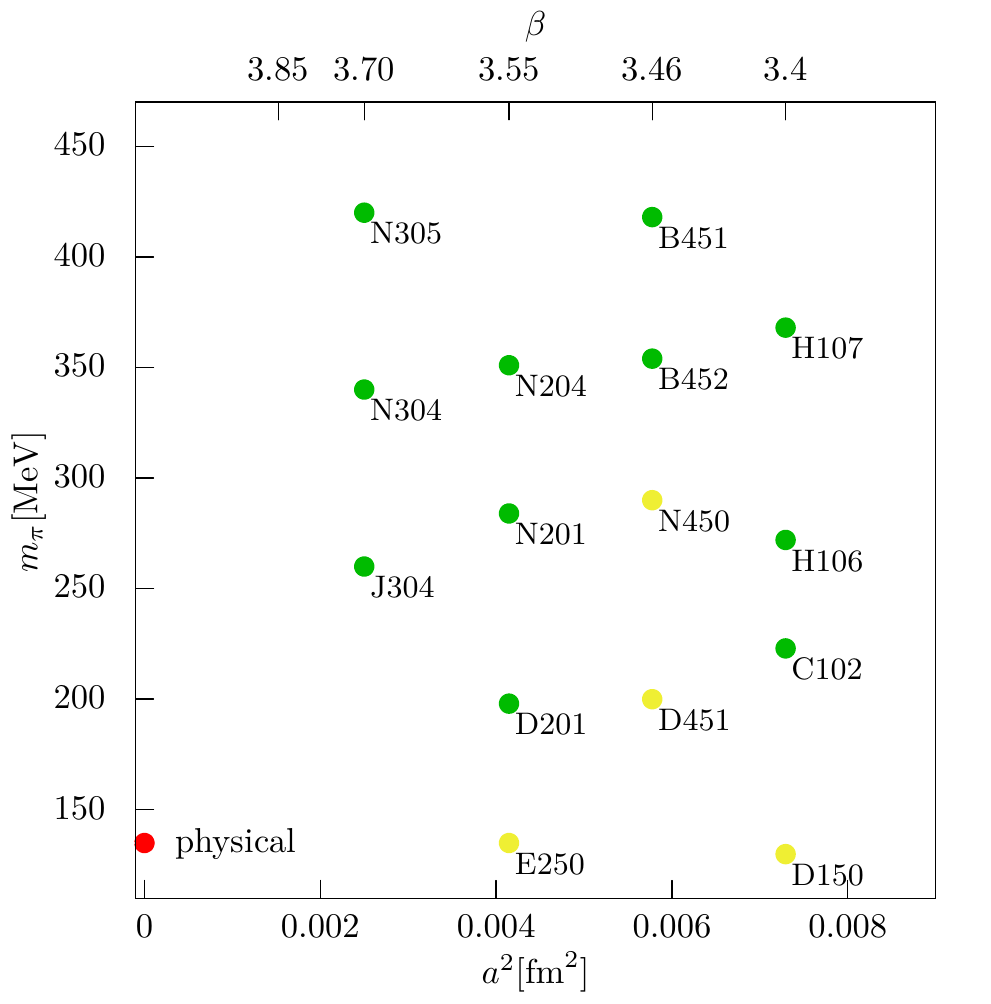}
\includegraphics[width=0.3\textwidth]{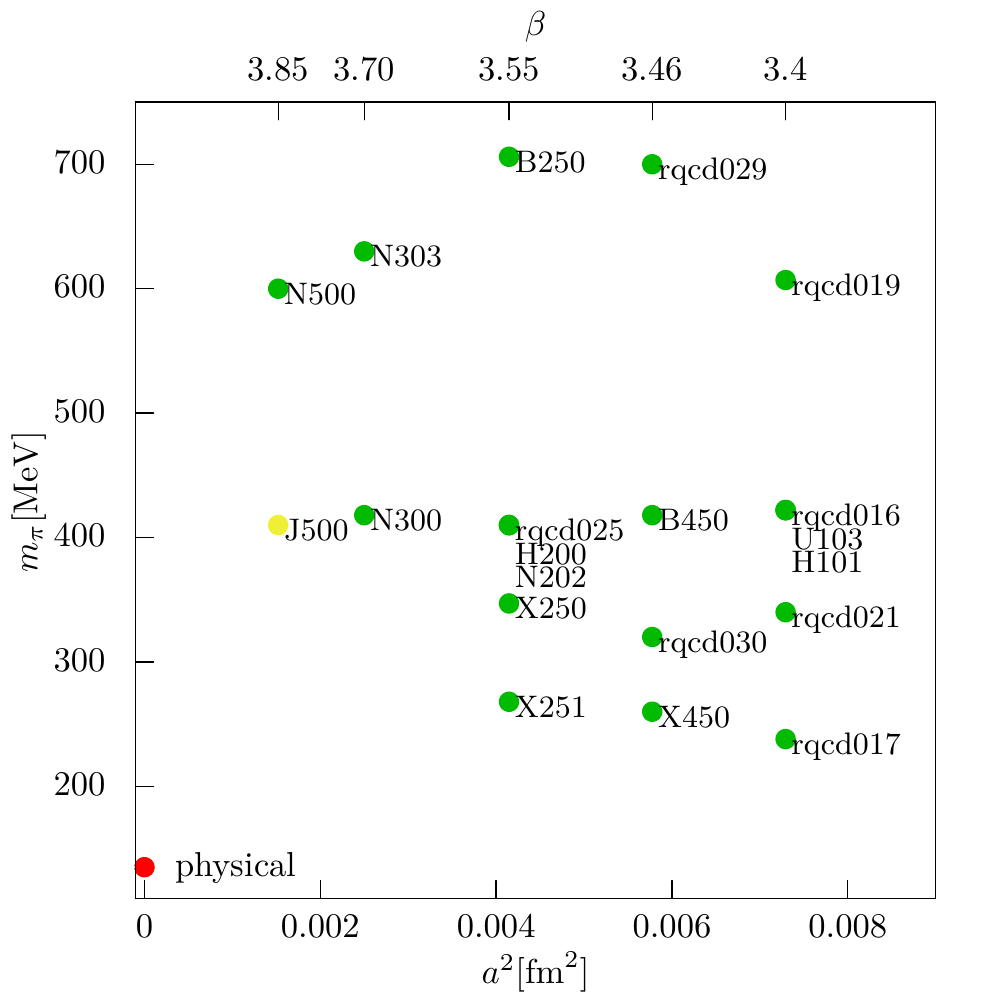}
\caption{Landscape of CLS ensembles as a function of the lattice spacing $a^2$ and pion mass $m_{\pi}$ along the three trajectories in the $m_l - m_s$ plane. Yellow ensembles are still in production. Not all lattices shown above are part of the present analysis. In our forthcoming publication we will give more detail on the ensembles. From left to right the constant quark mass sum, constant strange quark mass, symmetric trajectories are shown. \label{fig. cls}}
\end{center}
\end{figure}

In this project we employ the gauge field ensembles generated by the CLS consortium. Some of the ensembles are described in Refs. \cite{Bruno:2014jqa} and \cite{Bali:2016umi} and a scale setting was presented in Ref. \cite{Bruno:2016plf}. We take advantage of several features of these ensembles. They cover a wide range of lattice spacings, from 0.086 fm down to 0.039 fm, with 5 different values of the lattice spacing. The pion mass is varied from 450 MeV down to its physical value. These two facts and the availability of three different trajectories in the plane of light and strange quark masses (a symmetric line where $m_l = m_s$, a line where the sum of quark masses is constant $m_s + 2 m_l \approx \textrm{const}$ and a line where the strange quark mass is kept at its physical value) provide us with a firm handle on both the continuum and chiral extrapolations. The landscape of used ensembles in the plane of $a^2$ and $m_{\pi}$ is shown in figure \ref{fig. cls}.

\vspace{-1.0cm}

\section{Improvements}
\label{sec. improvements}

\begin{figure}
\begin{center}
\includegraphics[width=0.47\textwidth]{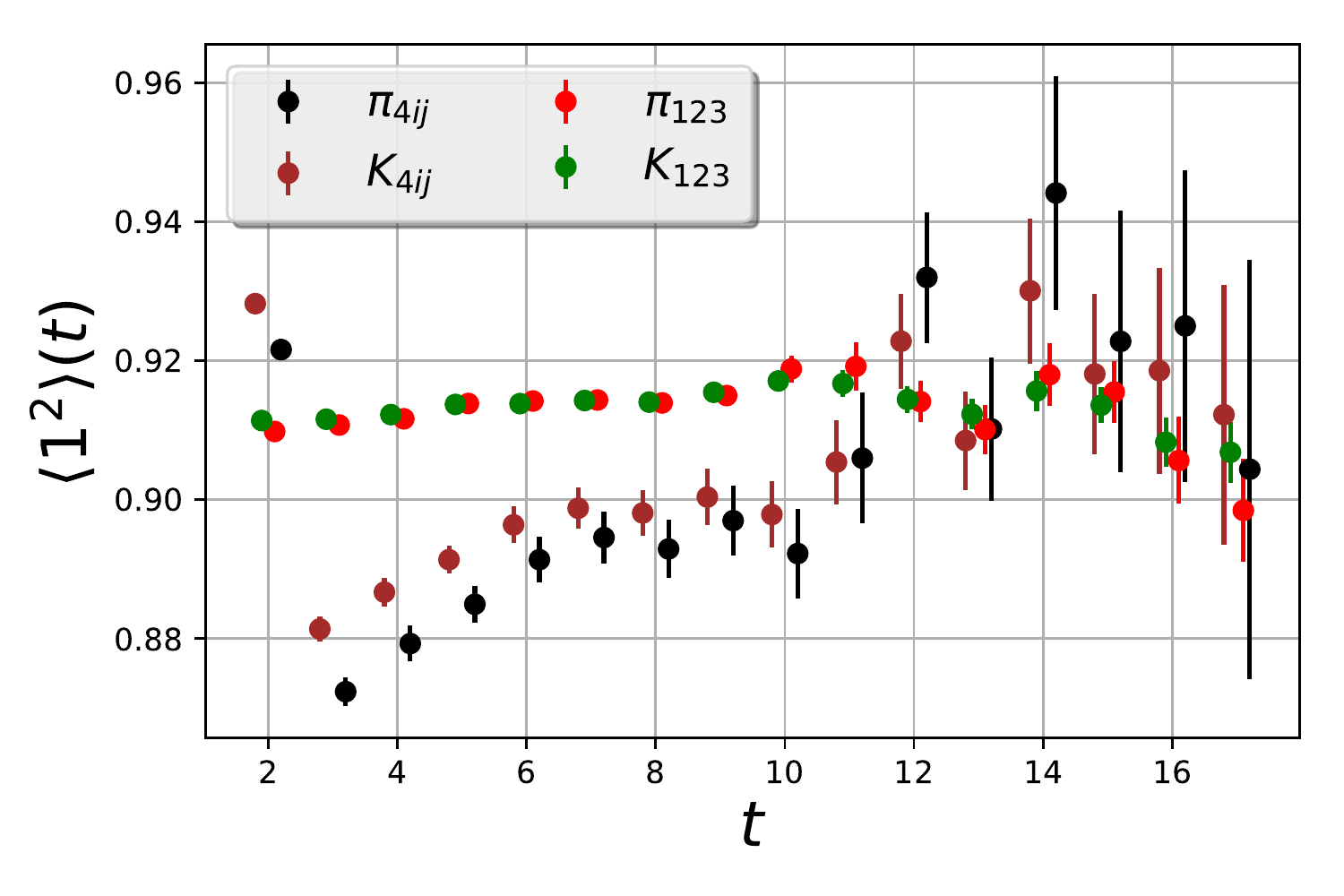}
\vspace{-0.5cm}
\caption{Comparison of signal for the correlation function $C^{+}_{\rho \mu \nu}(t,\mathbf{p})$ Eq.~\eqref{eq. correlation functions} as a function of time for different combinations of Lorentz indices. \label{fig. combination}}
\end{center}
\end{figure}

In the definition of the ratios $R^\pm_{\rho \mu \nu , \sigma}$ one has the freedom
to choose the indices $\mu$, $\nu$, $\rho$ and $\sigma$. In order to avoid complicated
mixing patterns the indices $\mu$, $\nu$, $\rho$ should all be different.
We are now employing a combination in which the derivative is never acting along the time direction, which exhibits a clear advantage in the signal quality and in terms of ground state overlap over all other combinations. As a demonstration we show in figure \ref{fig. combination} the data for different index combinations for the pion and kaon correlation functions $C^{+}_{\rho \mu \nu}(t,\mathbf{p})$ for the ensemble with $a=0.086$ fm and $m_{\pi} \approx 350$ MeV.

\section{Chiral and continuum extrapolations}
\label{sec. extrapolations}

The value of $\langle \xi^2 \rangle$ or $a_2$ in the continuum and for physical quark masses has to be obtained from an extrapolation from the measured data points. In this study we perform a combined fit to all data points (all lattice spacings and all pion/kaon masses along the three trajectories) using as an Ansatz a continuum ChPT formula which does not exhibit chiral logs \cite{Chen:2005js} supplemented with cutoff effects parametrization
\begin{align}
\langle \xi^2 \rangle_{\pi} &= \big( 1 + c_0 a + c_1 a \overline{M}^2 + c^{\pi}_{2} a \delta M^2 \big) \big(
 \langle \xi^2 \rangle_0 + \overline{A} \overline{M}^2 - 2 \ \delta A \ \delta M^2 \big),  \\
\langle \xi^2 \rangle_{K} &= \big( 1 + c_0 a + c_1 a \overline{M}^2 + c^{K}_{2} a \delta M^2 \big) \big(
 \langle \xi^2 \rangle_0 + \overline{A} \overline{M}^2 + \ \delta A \ \delta M^2 \big),
 \end{align}
with $\overline{A}$, $\delta A$ being low energy constants and
\begin{equation}
\overline{M}^2 = \frac{2m^2_K + m_{\pi}^2}{3}, \qquad \qquad \delta M^2 = 2 m^2_K - 2 m_{\pi}^2. 
\end{equation}
The fit is performed for both the pion and kaon second moment simultaneously. Altogether 7 parameters are fitted: the second moments in the chiral limit $\langle \xi^2 \rangle_0$, two low energy constants $\overline{A}$ and $\delta A$ and four coefficients parametrizing discretization effects $c_0$, $c_1$, $c_2^{\pi}$ and $c_2^{K}$.

\begin{figure}
\begin{center}
\includegraphics[width=0.47\textwidth]{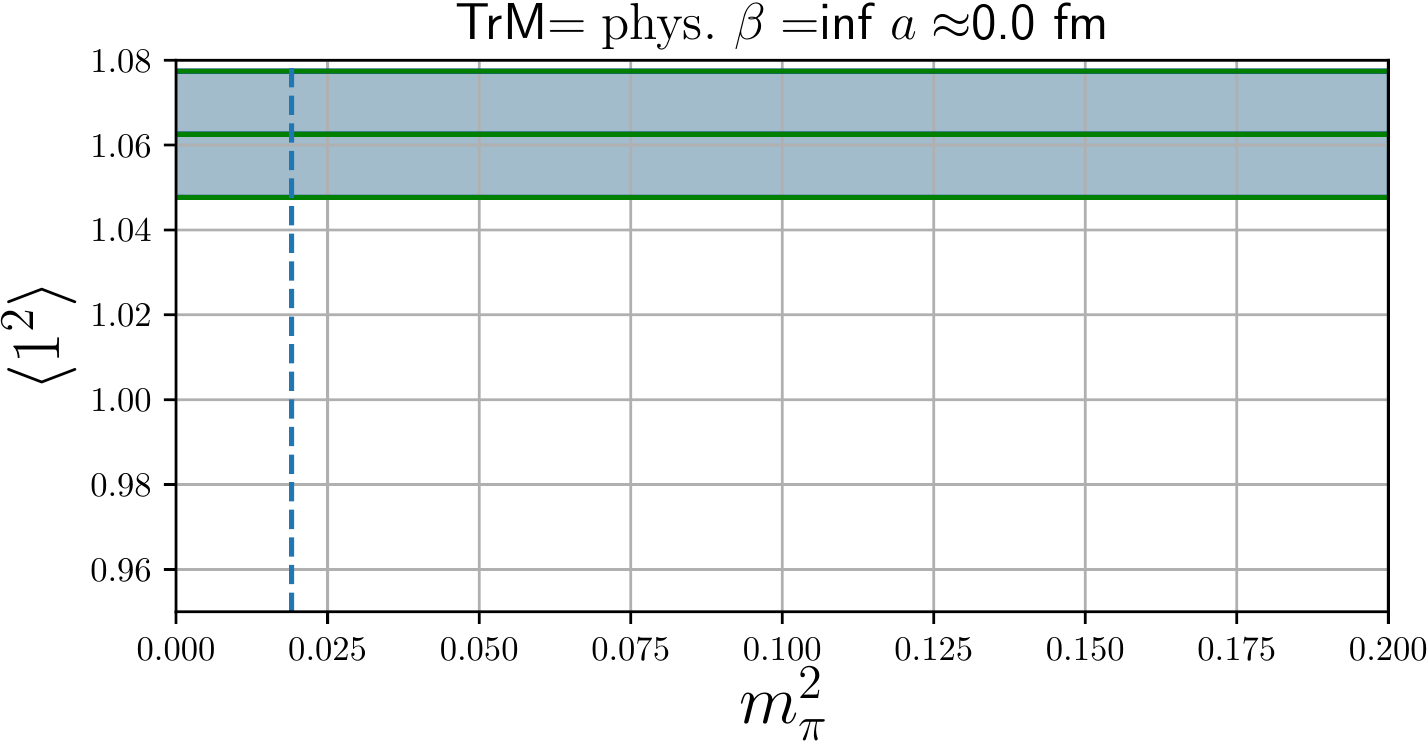}
\includegraphics[width=0.47\textwidth]{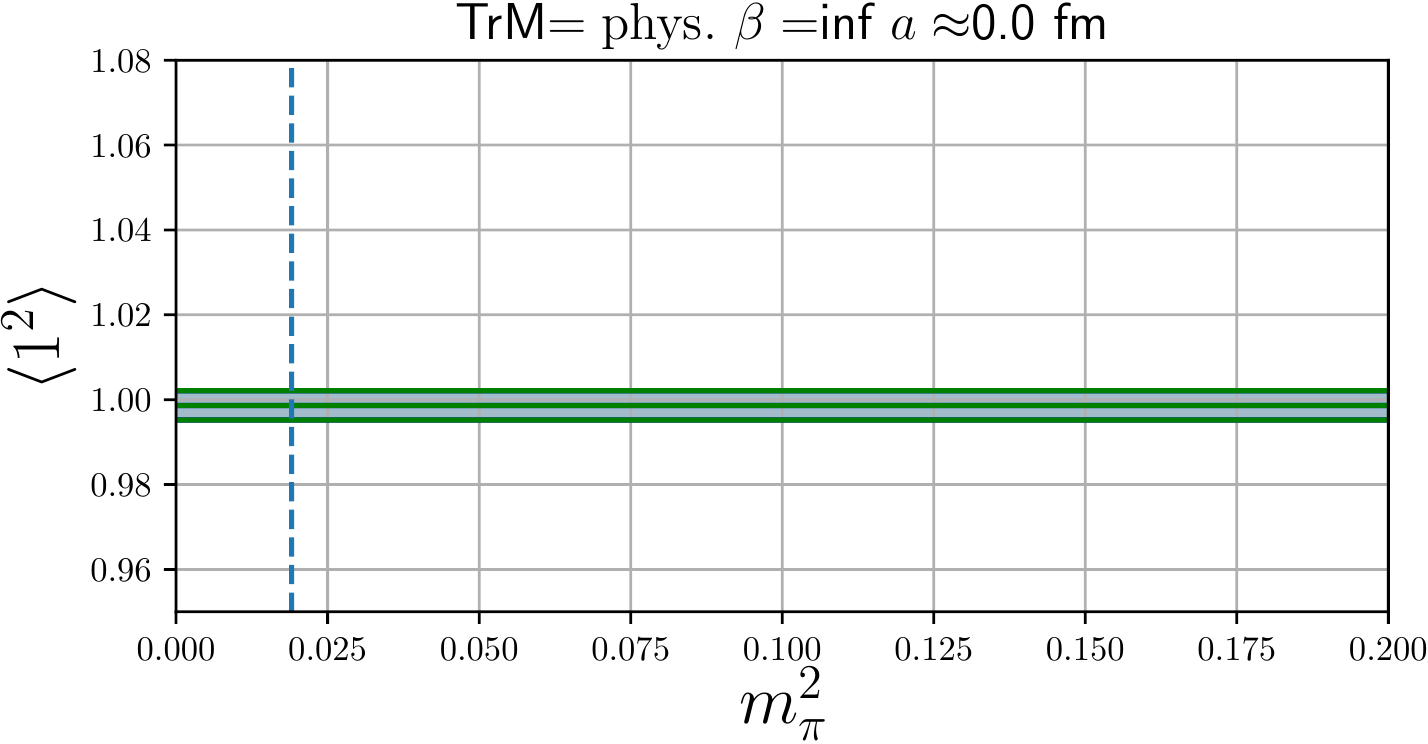}
\end{center}
\vspace{-0.5cm}
\caption{Continuum and chirally extrapolated value of the renormalized $\mathcal{O}^+$ operator Eq.~\eqref{eq. operators} for the old and new combination of Lorentz indices in the left and right panel respectively. \label{fig. unity}}
\end{figure}

As a check of the entire procedure we show the extrapolated value of the renormalized $\mathcal{O}^+$ operator which according to Eq.~\eqref{eq. operators} is equal to 1 in the continuum limit. However, the measured value can differ from 1 as can be already seen of the right panel of figure \ref{fig. combination} where the expectation value is approximately 0.91. We show the result of our fit in figure \ref{fig. unity}. In the left panel data obtained with the old set of Lorentz indices was extrapolated yielding a value significantly different from 1 with a large uncertainty. On the contrary, when we extrapolate the data where we exclude combinations when the derivative acts in the time direction, the extrapolated value is compatible with 1 and the uncertainty is considerably smaller. This result is a nice confirmation of the validity of the fitting Ansatz. 

\begin{figure}
\begin{center}
\includegraphics[width=0.47\textwidth]{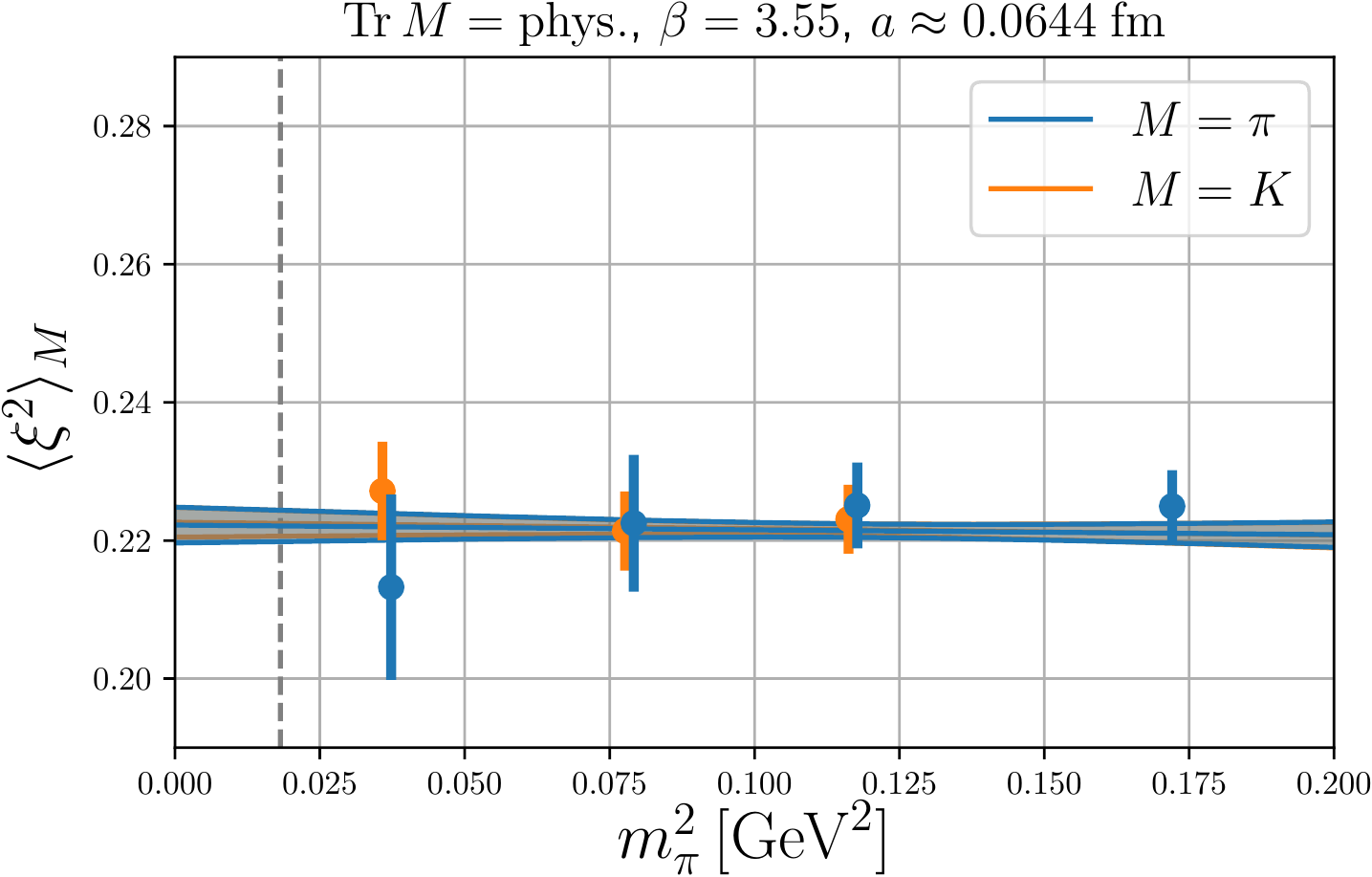}
\includegraphics[width=0.47\textwidth]{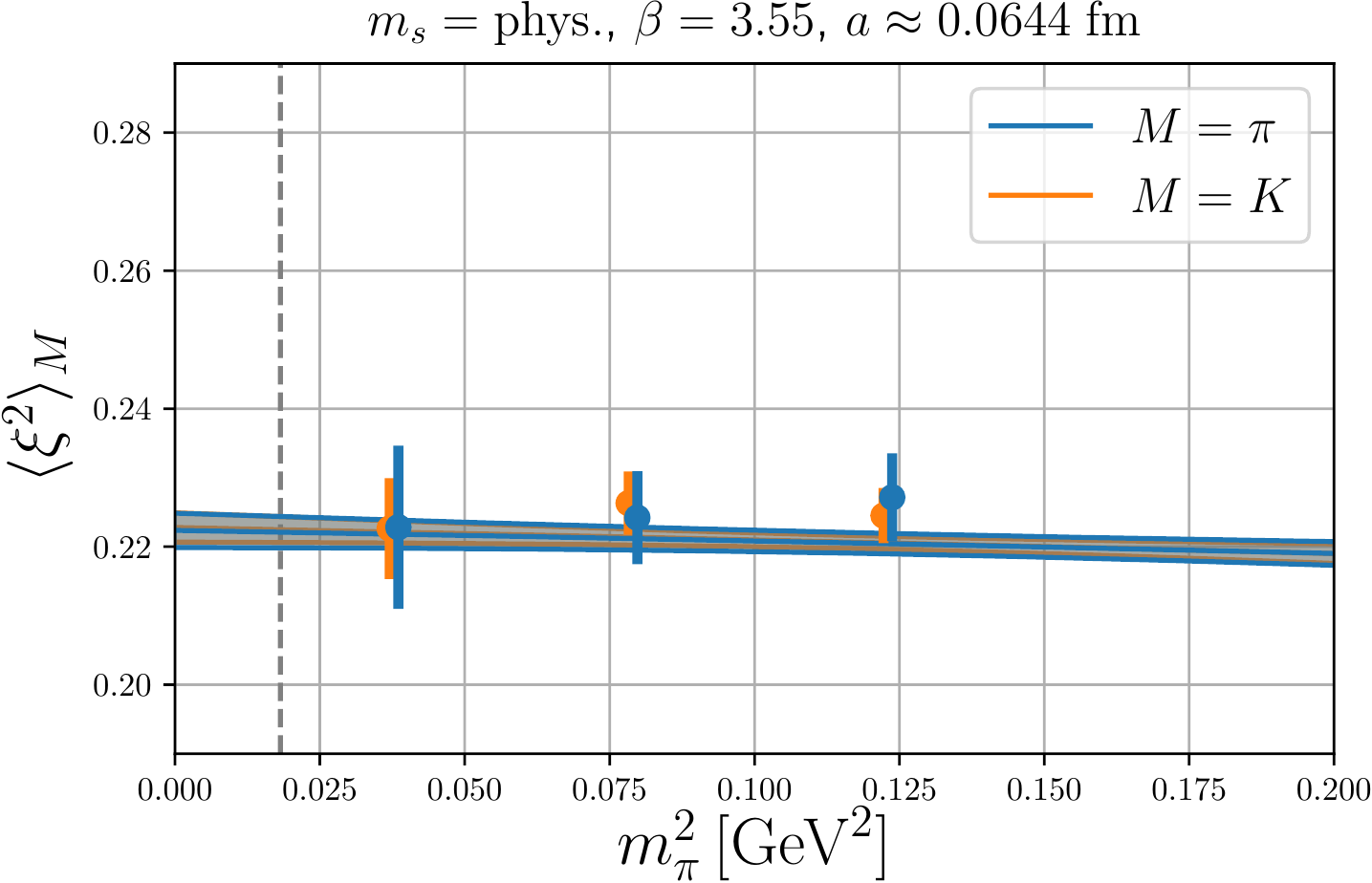}
\end{center}
\vspace{-0.5cm}
\caption{Extrapolation of $\langle \xi^2 \rangle$ along the constant trace (left panel) and constant strange quark mass (right panel) trajectories at $\beta=3.55$. \label{fig. extrapolation}}
\end{figure}

We apply the same extrapolation procedure to the combination of matrix elements yielding the second moments of the pion and kaon DAs and obtain their continuum values. Plot in figure \ref{fig. extrapolation} shows the extrapolation as a function of the pion mass along the constant sum of quark masses trajectory at $a=0.064$ fm. Our preliminary results are in the same ballpark as the previous estimates from $N_f=2$ simulations \cite{PhysRevD.92.014504}, however not all uncertainties have been carefully taken into account yet. We are completing our analysis using all the ensembles from the CLS landscape and the final results will be published soon.

\vspace{-0.5cm}

\section{Conclusions}
\label{sec. conclusions}

In this contribution we reported on the status of our project aiming at the non-perturbative determination of second moment of pion and kaon DAs in the continuum. We briefly discussed our framework and pointed out several improvements upon previous studies which allow us to control both the chiral and continuum extrapolations. As an example we showed the case of the $\langle 1^2 \rangle$ expectation value which gives a value in agreement with continuum expectations with a reasonable small uncertainty. We are now investigating systematic effects affecting our outcomes and final results should appear soon.

Finally, we would like to mention a recent advancement in the study of
hadron structure functions
which allows to circumvent the limitation of euclidean lattices and study
directly the x dependence
of structure functions from purely space-like correlation functions.
Several implementations of these ideas have already been
discussed in the context
of the pion DA.
The Authors of \cite{PhysRevD.95.094514} implemented  the Large Momentum Effective Theory
approach following the work of Ref. \cite{PhysRevLett.110.262002} and extracted the pion
DA directly in $x$-space. In
Ref. \cite{Bali:2017gfr} and \cite{Bali:2018spj} a different, coordinate space formulation was used \cite{Braun:2007wv}.
A comparison of the second moments obtained directly from the local
operator method as discussed in this
contribution and from these novel techniques will provide useful
crosschecks on the systematic uncertainties hidden in both
approaches.

\section*{Acknowledgements}
This work was supported by Deutsche Forschungsgemeinschaft under Grant No. SFB/TRR 55 and by the polish NCN grant No. UMO-2016/21/B/ ST2/01492.
We acknowledge the Interdisciplinary
Centre for Mathematical and Computational Modelling
(ICM) of the University of Warsaw for computer
time on Okeanos (grants No. GA67-12, GA69-20, GA71-26), the PLGRID consortium for computer time allocation on the Prometheus machine hosted by Cyfronet Krakow (grants hadronspectrum, nspt, pionda) and the Leibniz Rechenzentrum in Garching for access to the coolMUC3 cluster.
Additional computations have been carried out on the Regensburg QPACE~2 computer and the QPACE~3 machine of SFB/TRR~55 hosted by the Jülich Supercomputing Centre (JSC). The authors gratefully acknowledge the computing time granted by the John von Neumann Institute for Computing (NIC) and provided on the supercomputer JURECA-Booster at JSC.

\bibliographystyle{JHEP}
\bibliography{references}

%

\end{document}